\documentclass[useAMS,usenatbib]{mn2e}

\usepackage{graphicx}

\def\be{\begin{equation}} 

\def\ee{\end{equation}}

\def\gsim{\lower.5ex\hbox{\gtsima}} 

\def\lsim{\lower.5ex\hbox{\ltsima}} \def\gtsima{$\; \buildrel > \over 

\sim \;$} \def\ltsima{$\; \buildrel < \over \sim \;$} \def\prosima{$\; 

\buildrel \propto \over \sim \;$} \def\gsim{\lower.5ex\hbox{\gtsima}} 

\def\lsim{\lower.5ex\hbox{\ltsima}} 

\def\simgt{\lower.5ex\hbox{\gtsima}} 

\def\simlt{\lower.5ex\hbox{\ltsima}} 

\def\simpr{\lower.5ex\hbox{\prosima}} \def\la{\lsim} \def\ga{\gsim}

 \def\gtsima{$\; \buildrel > \over \sim \;$} 

\def\ltsima{$\; \buildrel < \over \sim \;$} 

\def\gsim{\lower.5ex\hbox{\gtsima}} 

\def\lsim{\lower.5ex\hbox{\ltsima}} 

\def\simgt{\lower.5ex\hbox{\gtsima}} 

\def\simlt{\lower.5ex\hbox{\ltsima}} 

\def\simpr{\lower.5ex\hbox{\prosima}} 

\def\la{\lsim} 

\def\ga{\gsim}

\def\E3{{\cal E}_{\rm g}^{III}}

\def\Msun{\rm M_\odot}

\def\Zsun{\rm Z_\odot}

\def\M*{M_*}

\def\Z*{Z_*}

\def\L*{L_*}

\def\MUV{M_{UV}}

\def\EBV{E(B-V)}

\title[LG progenitors amongst high-z LBGs]{CLUES to the past: Local Group progenitors amongst high-redshift Lyman Break Galaxies}
\author[Dayal et al.]{Pratika Dayal$^{1}$\thanks{E-mail:prd@roe.ac.uk (PD)}, Noam I. Libeskind $^{2}$ \& James S. Dunlop$^{1}$ \\ 
$^{{1}}$ SUPA\thanks{Scottish Universities Physics Alliance}, Institute for Astronomy, University of Edinburgh, Royal Observatory, Edinburgh, EH9 3HJ, UK \\
$^{2}$  Leibniz-Institute for Astrophysics, Potsdam, An der Sternwarte 16, Potsdam, Germany, 14482}

\begin{document} 
\date{} 
\pagerange{\pageref{firstpage}--\pageref{lastpage}} \pubyear{} 
\maketitle

\label{firstpage} 
\begin{abstract}
We use state-of-the-art numerical simulations to explore the observability and the expected physical properties of the progenitors of the local group galaxies at $z \simeq 6-8$, within 1 billion years of the big bang. We find that the most massive progenitors of the Milky Way (MW) and Andromeda (M31) at $z \simeq 6$ and $7$ are predicted to have absolute UV continuum magnitudes $\MUV \simeq -17$ to $-18$, suggesting that their analogues lie close to the detection limits of the deepest near-infrared surveys conducted to date (i.e. HST WFC3/IR UDF12). This in turn confirms that the majority of currently known $z \simeq 6 - 8$ galaxies are expected to be the seeds of present-day galaxies which are more massive than $\L*$ spirals. We also discuss the properties of the local-group progenitors at these early epochs, extending down to absolute magnitudes $\MUV \simeq -13$. The most massive MW/M31 progenitors at $z \simeq 7$ have stellar masses $\M* \simeq 10^{7.5-8} \Msun$, stellar metallicities $\Z* \sim 3-6\%\,\Zsun$, and predicted observed UV continuum slopes $\beta \simeq -2.4$ to $-2.5$.
\end{abstract} 



\section{Introduction} 

\label{intro}


In the last three years, the sensitive near-infrared imaging made possible by Wide Field Camera 3 (WFC3/IR) on the {\it Hubble Space Telescope} (HST) has transformed the study of galaxies at $z \simeq 7$ \citep[see][]{dunlop2012b}. Several hundred Lyman-break galaxies (LBGs) have now been uncovered with HST in the redshift range $6.5 < z < 8.5$ \citep[e.g.][]{oesch2010, bouwens2010a, mclure2010, finkelstein2010, mclure2011, lorenzoni2011, bouwens2011, oesch2012,mclure-udf2012}, and the new ground-based wide-field near-infrared surveys, such as UltraVISTA \citep{mccracken2012} are now reaching the depths required to reveal galaxies more luminous than $\MUV \simeq -19$ at $z \simeq 7$ \citep{castellano2010a, ouchi2010, bowler2012}.

Most recently, attention has been refocused on pushing to even fainter magnitudes and still higher redshifts through the ultra-deep WFC3/IR imaging in the HUDF (GO 12498; hereafter UDF12). The recently-completed 128-orbit UDF12 observations reach $5\sigma$ detection limits of $Y_{105} = 30.0$, $J_{125} = 29.5$, $J_{140} = 29.5$, $H_{160} = 29.5$ (after combination with the UDF09 data), and are the deepest near-infrared images ever taken \citep{ellis-udf2012, dunlop-udf2012, mclure-udf2012, schenker-udf2012, ono-udf2012}. A detailed description of the UDF12 data-set is provided by \citet{koekemoer2012}, and the final reduced images are available on the UDF12 team web-page \footnote{http://udf12.arizona.edu}.

The ever-improving data on galaxies at early cosmic times has motivated attempts to link these objects to the galaxies seen in the local Universe, and in particular to the Milky Way (MW) or indeed to $\L*$ galaxies in general (which dominate the local luminosity density). The approaches used to establish such a link range from the use of correlation functions to state-of-the-art cosmological simulations. For example, tracing merger trees using cosmological simulations, \citet{nagamine2002} found that the brightest LBGs (with $M_V \leq -23$) at $z =3$ merge into groups and clusters at $z=0$, and about half the LBGs with $-23 \leq M_V \leq -22$ evolve into local $\L*$ galaxies. Meanwhile, using a clustering analysis of a sample of 28500 galaxies at $z \simeq 1.4-3.5$, \citet{adelberger2005} concluded that the typical objects in their sample would have evolved into elliptical galaxies at $z =0$. More recently, coupling N-body Dark Matter (DM) simulations with a prescription for star formation, feedback and chemical evolution, \citet{okrochkov2010} have claimed that the progenitors of the MW could be visible as LBGs up to $z=7$ and $z=9$, including and excluding the effects of dust absorption on ultraviolet (UV) photons, respectively. The reader is referred to \citet{salvadori2010}, \citet{guaita2010}, \citet{dayal-libeskind2012}, \citet{walker-soler2012} and \citet{yajima2012b} for calculations aimed at establishing whether local group (LG) progenitors could be detected as Lyman Alpha Emitters (LAEs) in existing or future surveys.

In this study, our aim is to investigate the link between the faintest high-redshift LBGs uncovered to date (i.e. galaxies with $\MUV \simeq -17$\,AB mag detected through the UDF12 campaign) and local $\L*$ galaxies. Owing to computational limitations, at present, it is too costly to run a simulation with a large enough box size so as to be able to treat long-wavelength modes effectively, as well as a high enough spatial resolution to resolve galaxies as faint as those detected in the UDF12 imaging. We therefore employ the approach of using 
re-simulations of the LG: by construction, these reproduce the properties of the two local $\simeq \L*$ galaxies, the MW and M31 (Andromeda) at $z=0$, while the small volume simulated allows a high resolution to be achieved, as is discussed in more detail in what follows.



\section{The simulations}

\label{sims}


\begin{figure*} 
\center{\includegraphics[scale=0.6]{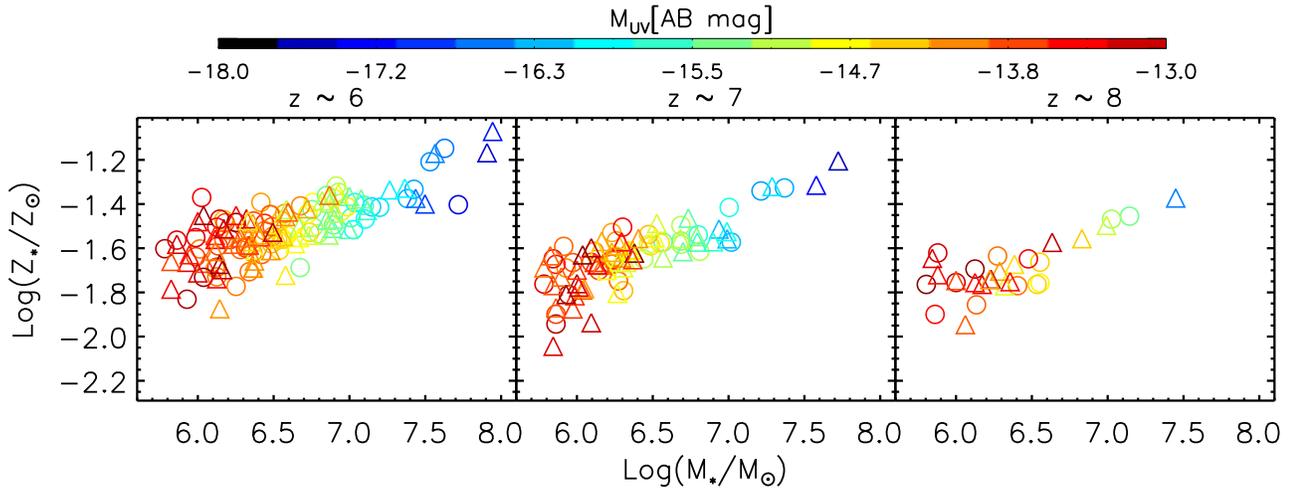}} 
 \caption{The mass-weighted stellar metallicity, $\Z*$, as a function of the stellar mass, $\M*$, for the simulated progenitor galaxies at $z \simeq 6-8$. Open circles and triangles show the $\Z*-\M*$ relation for the progenitors of the MW and M31 respectively. The points have been colour-coded according to the absolute UV magnitude shown on the colour-bar.  }
\label{massmet} 
\end{figure*}

For the calculations presented in this work, we have used the {\tt CLUES} simulation, which was run with the PMTree-SPH code \texttt{GADGET2} \citep{springel2005} in a cosmological box of size $64 h^{-1}$ comoving Mpc (cMpc). The runs used standard $\Lambda$CDM initial conditions and WMAP3 parameters \citep{spergel2007} such that $\Omega_m = 0.24$, $\Omega_{b} = 0.042$, $\Omega_{\Lambda} = 0.76$, $h = 0.73$, $\sigma_8 = 0.73$ and $n=0.95$. 

In brief, observations of objects in the local universe were used to reconstruct the initial density field using the Hoffman-Ribak algorithm \citep{hoffman-ribak1991}. These constrained initial conditions force the $z=0$ linear scales to resemble the input, the local universe; non-linear scales are unconstrained. To obtain a reliable LG, a number of low-resolution runs were performed until a suitable LG (defined in terms of the mass, relative distance and velocity of the two MW and M31 mass haloes) was found. These initial conditions were then re-run with high resolution and gas dynamics using the prescriptions given by \cite{klypin2001}, to produce two objects resembling the MW and M31, each of which contains about $10^6$ particles within their virial radii; interested readers are referred to \cite{libeskind2010}, \cite{knebe2010}, \cite{libeskind2011} and \cite{knebe2011} for complete details. Using a spatial resolution of 150~pc, the mass resolution obtained was $2.1\times10^{5}h^{-1} \Msun$ and $4.4\times10^{4}h^{-1} \Msun$ for DM and gas particles, respectively. 

The simulation employs the feedback rules of \cite{springel-hernquist2003b}: hot ambient gas and cold gas clouds in pressure equilibrium form the two components of the interstellar medium (ISM). Gas properties are calculated assuming a uniform but evolving UV background generated by QSOs and AGNs \citep{haardt-madau1996}. Metal line (or molecular) dependent cooling below $10^{4}{\rm K}$ is ignored. Star formation is treated stochastically, choosing model parameters that reproduce the Kennicutt law for spiral galaxies \citep{kennicutt1983, kennicutt1998}. The instantaneous recycling approximation is assumed: cold gas cloud formation (by thermal instability), star formation, the evaporation of gas clouds, and the heating of ambient gas by supernova driven winds all occur at the same instant. We assume kinetic feedback in the form of winds driven by stellar explosions.

Haloes and subhaloes are identified using the publicly available Amiga Halo Finder \citep[AHF;][]{knollmann-knebe2009}. AHF uses an adaptive grid to find local over-densities; after calculating the gravitational potential, particles that are bound to the potential are assigned to the halo. 

In order to find the progenitors of the MW and M31 that could be visible as LBGs at $z\simeq 6-8$, we identify all the particles within their virial radii at $z=0$. These particles are then followed back in time and located at $z\simeq 6, 7$ and 8 in the simulation box; if bound to a structure, the corresponding halo is considered to be the progenitor of a local $\L*$ galaxy at that redshift. Further, in all the calculations presented in this paper, we only use those high-redshift progenitors of the MW and M31 that contain at least 20 star particles; this corresponds to galaxies that contain a minimum of 540 total (DM, gas and star) particles at $z \simeq 6-8$.

Although we have used an extremely high-resolution simulation that reproduces the masses, relative velocities and separation of the two local $\L*$ galaxies (the MW and M31) to trace their progenitors back in time, the main limitation of this work is that all the results shown in this paper relate to the specific merger tree obtained from the simulation. It would be interesting to explore how these results might change in the future when/if a statistically-significant number of similar resolution LG simulations become available.


\subsection{Identifying LBGs}

\label{find_lbgs}


To identify the progenitors of the MW and M31 that could be observed as high-redshift LBGs, we start by computing their UV luminosities (or absolute magnitudes): we consider each star particle to form in a burst, after which it evolves passively. The total spectral energy distribution (SED), including both the stellar and nebular continuum, for each star particle is computed via the population synthesis code {\tt STARBURST99} \citep{leitherer1999}, using its mass, stellar metallicity and age. The age is computed as the time difference between the redshift of the snapshot and the redshift of formation of the star particle. Further, we assume a zero escape fraction of neutral hydrogen ionizing photons, thereby maximizing the contribution of the nebular continuum to the SED. The total intrinsic continuum luminosity, $L_c^{int}$ (at 1500 \AA\, in the galaxy rest frame) for each progenitor is then calculated by summing the SEDs of all its star particles. 

Continuum photons can be absorbed by dust within the ISM and only a fraction, $f_c$, escape out of any progenitor, unattenuated by dust. The value of $f_c$ for each progenitor is calculated assuming Type II supernovae (SNII) to be the primary dust factories such that each SNII produces about $0.4 \Msun$ of dust \citep{todini-ferrara2001}, each SNII destroys dust with an efficiency of about 20\% within the region it shocks to speeds $\geq 100~{\rm km \, s^{-1}}$ \citep{seab-shull1983}, a homogeneous mixture of gas and dust is astrated into star formation, and that dust is lost in SNII-powered outflows. We assume the dust to be distributed in a screen of radius $r_d = r_g$ where the gas distribution radius is calculated to be $r_g = 4.5\lambda r_{200}$. Here, the spin parameter has a value $\lambda=0.05$ \citep{ferrara2000}, and $r_{200}$ is the virial radius. This dust distribution radius is used to obtain the dust surface mass density which can then be translated in the dust optical depth to obtain $f_c$ \citep[for details of this calculation see][]{dayal2010a, dayal2013}. The observed continuum luminosity $L_c^{obs}$ can then be expressed as $L_c^{obs} =L_c^{int} \times f_c$.  

\section{Results}
\label{results}

As a result of the exquisite simulation resolution we are able to study LBGs that are nearly two orders of magnitude fainter in luminosity ($\MUV < -13$) than the current near-infrared detection limit of $\MUV \simeq -17$ at $z\simeq 7$; such faint LBGs could potentially be detected with future facilities such as the {\it James Webb Space Telescope} (JWST). We now present the theoretical results for the physical properties of the MW and M31 progenitors, and the resulting predictions for observables such as the luminosity functions (LFs) and UV spectral slopes, $\beta$ ($f_{\lambda} \propto \lambda^{\beta}$), for comparison with the most recent observational results presented by \citet{mclure-udf2012} and \citet{dunlop-udf2012}.

\begin{figure*} 
\center{\includegraphics[scale=1.0]{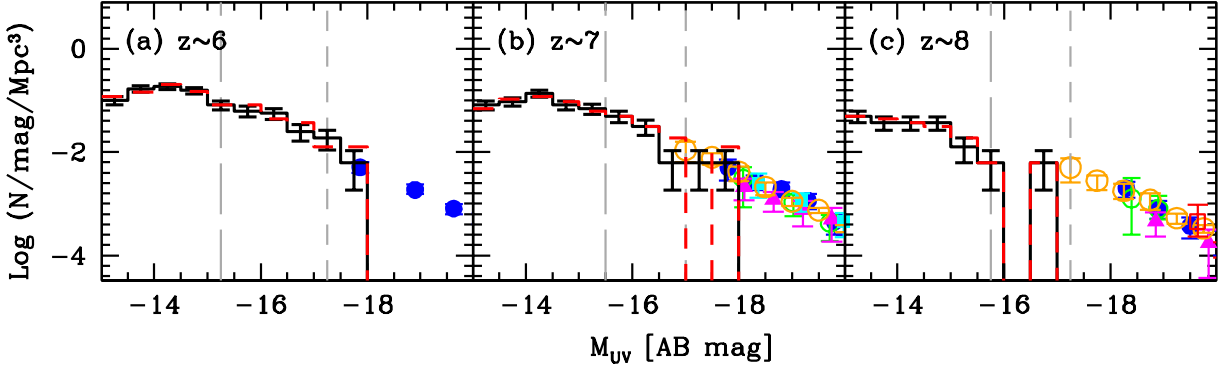}} 
\caption{The UV LFs of the progenitors of the MW and M31 at $z \simeq 6-8$. In each panel, points show the observational results, while the dashed (solid) histograms show the intrinsic (dust attenuated) UV LFs obtained using the progenitors of both the MW and M31. The observed LBG UV LFs have been taken from: (a) $z\simeq 6$: \citet[ blue filled circles]{bouwens2007}; (b) $z \simeq 7$: \citet[cyan filled squares]{oesch2010}, \citet[green empty circles]{bouwens2010}, \citet[blue filled circles]{bouwens2011}, \citet[red empty triangles]{castellano2010a}, \citet[magenta filled triangles]{mclure2010} and \citet[orange empty circles]{mclure-udf2012}; (c) $z \simeq 8$: \citet[green empty circles]{bouwens2010}, \citet[blue filled circles]{bouwens2011}, \citet[magenta filled triangles]{mclure2010}, \citet[red empty squares]{bradley2012} and \citet[orange empty circles]{mclure-udf2012}. The vertical short (long) dashed lines in each panel show the detection limits for HST (JWST) near-infrared imaging.}
\label{uvlf} 
\end{figure*}

\subsection{The mass-metallicity-magnitude relation}
\label{sec_mmm}

We start by showing the relation between the stellar mass ($\M*$), mass-weighted stellar metallicity ($\Z*$) and the dust-attenuated absolute UV magnitude ($\MUV$) for the progenitors of the MW and M31, in Fig. \ref{massmet}. As expected from the hierarchical structure formation scenario where successively larger systems assemble from the merger of smaller ones with time, the $\M*$ range increases with decreasing redshift, going from $10^{5.7-7.5}\, \Msun$ to $10^{5.7-8.0}\, \Msun$ in the 300 Myr between $z \simeq 8$ and $z \simeq 6$. This is accompanied by a significant increase in the number of identifiable MW and M31 progenitors, from about 64 at $z \simeq 8$ to 210 at $z \simeq 6$, as more and more tiny building blocks assemble with time and undergo star formation. Since metals are produced by star formation, progenitors that have built up a larger stellar mass are also more metal rich, giving rise to a mass-metallicity relation such that $\Z*$ increases from $10^{-1.9}$\,$\Zsun$ for $\M* = 10^{5.7} \Msun$ to about $10^{-1.08}$\,$\Zsun$ for $\M* = 10^{8.0} \Msun$ at $z \simeq 6$; the average stellar metallicity rises from about 0.02\,$\Zsun$ at $z \simeq 8$ to 0.035\,$\Zsun$ at $z \simeq 6$. As a result of their larger potential wells (that prevent gas loss due to negative feedback from SN-driven winds), more massive galaxies contain larger gas masses available for star formation, and hence are the brightest in the rest-frame UV. 

To summarize, we find a three-dimensional relation between $\M*-\Z*-\MUV$ such that the more massive progenitors are both more metal enriched and brighter in the UV; these trends hold true for the progenitors of both the MW and M31. However, since the simulated stellar mass of M31 ($1.98 \times 10^{10} \Msun$) is larger than that of the MW ($1.62 \times 10^{10} \Msun$) at $z=0$, the most massive progenitors belong to M31 at any of the redshifts studied. 

Because a few of the objects shown in Fig. 1 have $\MUV\leq -17$, we would predict that analogues of some of the most massive progenitors of the MW and M31 at $z \simeq 6$, and of M31 at $z \simeq 7$ are bright enough to be found among the faintest high-redshift LBGs detected by the HST at these redshifts.  However, at $z \simeq 8$, {\it none} of the progenitors of the two $\L*$ galaxies at $z=0$ are brighter than the current detection limit of $\MUV<-17.25$, as is also shown in Fig. \ref{uvlf} that follows. 

\subsection{UV luminosity functions}
\label{sec_uvlf}

In Fig. \ref{uvlf} we compare the predicted UV LF of the progenitors of the MW and M31 to the observed UV LF at $z =6-8$. Within the re-simulated region, we find the centre of mass of the MW and M31 and build spheres that enclose 95\% of the total mass assembled at that redshift: the combined volume occupied by the progenitors of the MW and M31 is then found to be $\approx (120,200,175)\, {\rm cMpc^3}$ at $z \simeq (6,7,8)$. However, to estimate the {\it contribution} of the LF of the progenitors to the overall LF in a representative volume, we need to account for the typical comoving volume occupied by galaxies such as the MW and M31.

At $z=0$, the MW and M31 have absolute $K$-band magnitudes of $M_K= -24.02$ and $-24.51$ \citep{hammer2007}. To estimate the present-day number density of such galaxies we use the $K$-band LF of \citet{kochanek2001} after correcting their results to the Hubble constant adopted in our simulation; after boosting the absolute magnitudes given in Table 2 of \citet{kochanek2001} by 0.68 mag, and reducing the values of $\log n$ by $-0.41$ (i.e. to correct to $h= 0.73$), we find that the number density of galaxies in the absolute magnitude range $-24.7 < M_K < -23.7$ is $\simeq 6.2 \times 10^{-3} {\rm cMpc^{-3} \, mag^{-1}}$. This implies that one $\L*$ galaxy at $z=0$ typically occupies a volume of about $160 \,{\rm cMpc^3}$, implying that the total volume occupied by the MW and M31 is approximately $320 \,{\rm cMpc^3}$. Normalizing the combined volume occupied by the progenitors of both the MW and M31 at $z \simeq 6,7$ and $8$ to the volume of $320 \,{\rm cMpc^3}$ occupied by two typical $\L*$ galaxies at $z=0$, we obtain the UV LFs shown in Fig. \ref{uvlf}. 

As seen from the same figure, at $z \simeq 6$ and $z \simeq 7$, both including/excluding the effects of dust on UV photons, the most luminous progenitors of the MW and M31 are bright enough to be amongst the faintest LBGs that have been detected to date. Furthermore, in the small range of overlap with the data at these redshifts, the slope and the amplitude of the theoretical UV LF is in reasonable agreement with the observations; this agreement is particularly striking at $z \simeq 7$, where the deepest data are available, given that once the $f_c$ values have been calculated for any galaxy, there are no free parameters left to calculate the UV LF. However, at $z \simeq 8$, even the predicted intrinsic UV LF of the MW/M31 progenitors falls short of the current detection limits. 

Hence, our simulation suggests that, with the current detection limits, the most massive progenitors of local $\L*$ galaxies are luminous enough to only be visible at $z \leq 7$; above this redshift, the progenitors of local $\L*$ galaxies are too small to support enough star formation to be visible in the UV within the UDF12 data. Clearly our results thus also imply that high-redshift LBGs with $\MUV<-18$ are likely to be the progenitors of systems more massive than $\L*$ galaxies at $z=0$. Finally, since the number of MW and M31 progenitors increases dramatically with decreasing luminosity, it is clear that many such objects should be detectable with JWST, even at $z \simeq 8$.

\subsection{UV Spectral Slopes}
\label{sec_beta}
\begin{figure*} 
\center{\includegraphics[scale=1.0]{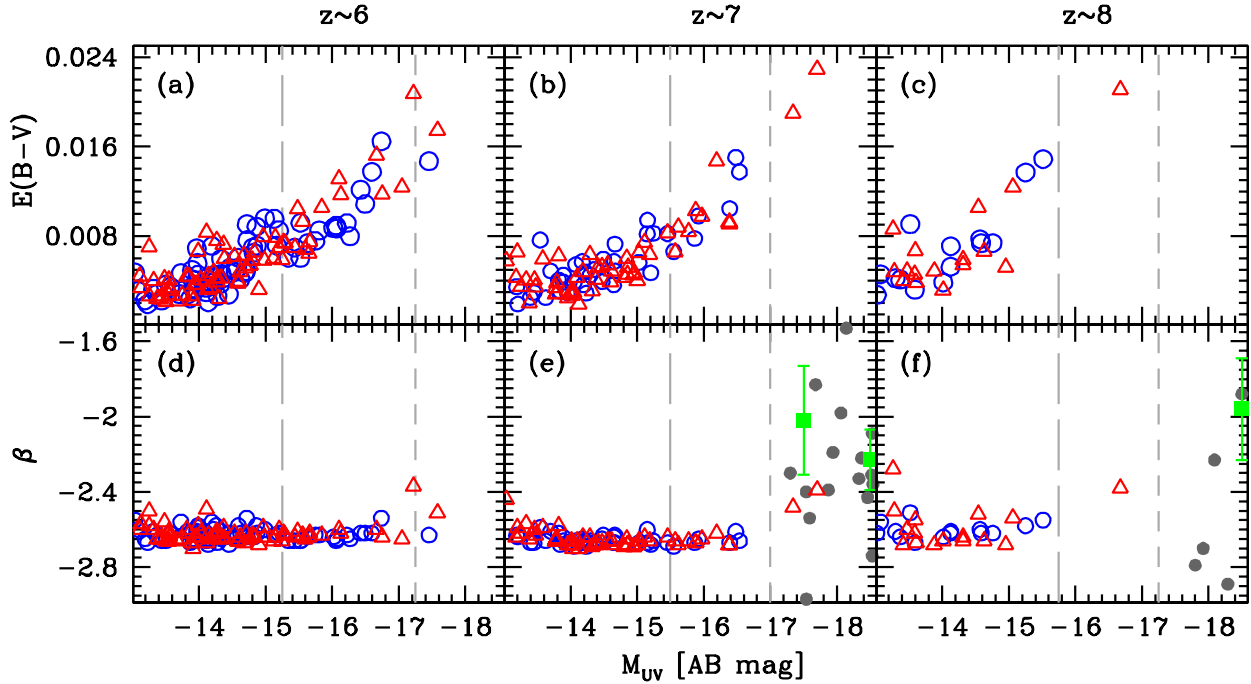} }
\caption{The upper and lower panels show the colour excess, $\EBV$, and the predicted UV spectral slope, $\beta$, as a function of absolute UV magnitude for the progenitors of the MW (empty circles) and M31 (empty triangles) for $z \simeq 6-8$ as marked above each panel. The vertical short (long) dashed lines show the detection limits for the HST (JWST). In panels (e) and (f), filled circles show the $\beta$ slopes for individual galaxies, and filled squares show the average beta values in bins of one magnitude with the associated errors, inferred observationally by \citet{dunlop-udf2012}. } 
\label{beta} 
\end{figure*}

We now explore the predicted colour excess, $\EBV = -2.5 {\rm log}(f_c)/11.082$, produced by the dust attenuation in the simulated galaxies. We have used the SN extinction curve \citep{bianchi-schneider2007} to infer the colour excess from the escape fraction of continuum photons for consistency with our assumption that SNII are the main dust producers in these early galaxies. Further, the SN extinction curve has also been shown to successfully explain the observed properties of quasars at $z \simeq 6$ \citep{maiolino2006} and gamma-ray bursts at $z =6.3$ \citep{stratta2007}. 

Since we assume SNII to be the main dust producers, and the SNII rate is related to the total stellar mass that has been formed in stars, the more massive a galaxy is in $\M*$, the larger its dust mass (and hence attenuation) is expected to be. This is exactly the trend seen from the upper 3 panels of Fig. \ref{beta} where the brightest progenitors show the largest colour excess such that $\EBV \simeq 0.02$ while the faintest systems show negligibly-small values of $\EBV \simeq 0.001$ at all the redshifts considered. The average value of $\EBV$ increases from about 0.004 at $z \simeq 8$ to 0.006 at $z \simeq 6$, as progenitors increase in mass and build up their dust content. 

To obtain the observed spectral slopes, $\beta$, for the progenitors of the MW and M31, we apply the SN extinction curve to the intrinsic spectrum of each progenitor, scaled according to the value of $f_c$ calculated for that progenitor. We use 30 values of $\beta$ evenly spaced between $-3$ and $0$ to fit a line through the observed SED of each galaxy (as obtained using {\tt STARBURST99}) over the wavelength range $\lambda_{rest} = 1500-3000$\AA\, sampled every 100 \AA; the value of $\beta$ yielding the minimum $\chi^2$ error is then chosen as the $\beta$ value for that particular galaxy. We find that $\beta$ increases slightly with increasing mass at any given redshift as shown in the lower three panels of Fig. \ref{beta}: for example, $\beta$ increases from about $-2.64$ to $-2.35$ as $\MUV$ increases from $-13.0$ to $\simeq -17.5$ at $z \simeq 6$; this is due to increasing dust masses and attenuation values with increasing $\M*$ as discussed above. 

Including the inferred amounts of dust attenuation makes the observed $\beta$ slightly redder than the intrinsic values, such that the average value of $\beta$ is $\langle \beta \rangle \approx (-2.54,-2.58,-2.6)$ for $z \approx (6,7,8)$, as compared to an average intrinsic $\beta$ value of $\langle \beta \rangle -2.64$ at any of the redshifts considered. However, this relatively small change in average $\beta$ produced by dust is dominated by the very small amounts of dust attenuation in the large number of fainter progenitors. What is most striking in Fig. 3 is that the currently-detectable galaxies at these redshifts are predicted to have much more significant dust reddening, resulting in typical individual values $\beta \simeq -2.4$. While this is still slightly bluer than the average value of $\beta = -2.02 \pm 0.29 \,(-1.96 \pm 0.27)$ for $M_{UV} \simeq -18$ at $z = 7 \, (8)$ recently derived by \citet{dunlop-udf2012}, it is clear that the observable values predicted here lie well within the range of $\beta = -4$ to $1$ ($-3.5$ to $-0.8$) observed for individual galaxies at $z \simeq 7 \,(8)$ as shown in panels (e) and (f) of Fig. \ref{beta} \citep[see also][]{dunlop2012}. Moreover, since the volume sampled by the UDF12 at $z \simeq 6.5 - 7.5$ is over an order of magnitude greater than that sampled here, the real galaxy samples at $z \simeq 7$ and $8$ inevitably contain many more massive galaxies than the most massive MW and M31 progenitors, and so the resulting sample-average values of $\beta$ would be expected to be higher (i.e. redder) than the UV slopes of even the most massive MW and M31 progenitors.

\section{Summary and Conclusions}
\label{conc}
We have coupled state-of-the-art high-resolution gas-dynamical simulations of the local group run within the \texttt{CLUES} framework (that reproduce the mass, relative distance and velocity of the two MW and M31 haloes), with a dust model to explore the predicted observability and physical properties of the progenitors of the MW and M31 at $z \simeq 6,7$ and $8$ down to absolute UV magnitudes $\MUV <-13$. Our main results can be summarized as follows.

\begin{itemize}

\item At $z \simeq 6,7$ and $8$, the simulated progenitors of the two $\L*$ galaxies at $z=0$ exhibit a three-dimensional relation between $\M*-\Z*-\MUV$, with the most massive progenitors being both the brightest in the UV, as well as the most metal rich.

\item The bright end of the predicted progenitor UV LFs at $z \simeq 6 \, (7)$ overlaps with the observed LFs for $\MUV \simeq -17.25$ to $-18.0$ ($-17.0$ to $-18.0$). In this small range of overlap, both the slope and the amplitudes of the theoretical and observed LFs are in excellent agreement. However, at $z \simeq 8$, even the intrinsic UV LF of the LG progenitors lies faint-ward of the observational limit at $\MUV \simeq -17.25$.

\item The above two results lead to a picture wherein at $z \simeq 6\, (7)$, analogues of the most massive progenitors of the MW (M31) are bright enough to be amongst the faintest LBGs that have been detected so far at these redshifts with $\MUV \simeq -17.25$ to $-18.0$ ($-17.0$ to $-18.0$). This implies that LBGs with $\MUV <-18$ at $z \simeq 6$ and $7$ are likely the progenitors of systems more massive than $\L*$ galaxies at $z=0$. At $z \simeq 8$, even the most massive progenitors of the MW and M31 are intrinsically fainter than the current limit of $\MUV =-17.25$ implying that all LBGs observed so far at this redshift are the progenitors of super-$\L*$ galaxies at $z=0$.

\item As a result of their relatively low stellar masses values at $z \simeq 6$ ($\M*  = 10^{5.7-8.0} \Msun$), and even lower masses at higher redshift, the vast majority of the progenitors of the two LG galaxies have tiny colour excess values ranging from $\EBV = 0.001$ to 0.02. The average $\EBV$ increases slightly from 0.004 at $z \simeq 8$ to 0.006 at $z \simeq 6$ as these galaxies build up their stellar and dust masses.

\item The dust attenuated UV spectral slopes become significantly redder with decreasing magnitude, from $\beta \simeq -2.64$ to $-2.35$ for $\MUV \simeq -13$ to $-17.5$ at $z \simeq 6$ as a result of the larger dust masses in the most massive (luminous) galaxies. The brightest progenitors at $z \simeq 7$ with $\MUV = -17.4$ and $-17.8$ exhibit $\beta$ values of $-2.5$ and $-2.4$ respectively. These values are slightly bluer than the average value of $\beta = -2.02 \pm 0.29$ inferred for $\MUV =-17.5$ LBGs at the same redshift by \citet{dunlop-udf2012}, but lie well within the observed scatter that ranges between $\beta = -4$ to $1$. This is broadly as expected, given that the most massive progenitors of the MW and M31 at these redshifts are comparable to the least luminous galaxies in the UDF12 samples studied by \citet{dunlop-udf2012}.
\end{itemize}


\section{Acknowledgements}

JSD and PD acknowledge the support of the European Research Council via the award of an Advanced Grant, and JSD also acknowledges the support of the Royal Society via a Wolfson Research Merit award. NIL is supported through a grant from the Deutsche Forschungs Gemeinschaft. The simulations were performed and analyzed at the Leibniz Rechenzentrum Munich (LRZ), the Neumann Institute for Computing (NIC) Juelich and at the Barcelona Supercomputing Centre (BSC). PD thanks E.J. Bernard for his invaluable help with the figures shown. Finally, the authors thank the anonymous referee for their extremely constructive comments.

\bibliographystyle{mn2e}

\bibliography{prog}

\newpage 

\label{lastpage} 

\end{document}